\documentstyle[12pt]{article}

\newcommand{\be}{\begin{equation}}
\newcommand{\ee}{\end{equation}}

\begin{document}

\begin{center}

{\Large {\bf Simulations of Single Polymer Chains in the Dense Limit}

\vspace{1.cm}}

{\bf Peter Grassberger and Rainer Hegger}

Physics Department, University of Wuppertal D-5600 Wuppertal 1, FRG

\vspace{.8cm}

\today

\vspace{1.1cm}

{\bf Abstract} \end{center}

{\small \advance \baselineskip by -2pt

We present simulation results for single a-thermal chain polymers
in finite volumes. For this we use a recently proposed recursive
implementation of the enrichment method. In 3 dimensions it allows
the simulation of extremely long chains (up to $N=300,000$). It is
much less efficient for $d=2$, but we can also there extend
considerably the previously accessible range of chain lengths and
densities. We verify all tested  scaling laws except one,
and we point out similarities with complex optimization problems.
}

\vspace{4.cm}


\eject

\section{Introduction}

Consider an ensemble of self avoiding chains (``SAW's") on a regular
lattice in
$d$ dimensions. All chains start from the origin of the lattice. Each
configuration of a chain of length $N$ (here and in the following, $N$
will be the number of links, so that the number of sites touched by the
chain is $N+1$) contributes to the partition function with a weight
$z^N$, where $z$ is a constant ($N$-independent) ``fugacity". If we
denote by $C_N$ the number of configurations with length $N$ and with
one end fixed at the origin (with $C_0=1$), we have thus
\be
    Z(z) = \sum_{N=0}^\infty z^N C_N  \;    \label{Z}.
\ee
This provides a simple model for the grand canonical
\footnote{Notice that some authors \cite{schaefer} call this
the {\it equilibrium ensemble}, in order to distinguish it from a truly
grand canonical ensemble in which also the number of chains can fluctuate;
in the present case, we always work with a single chain.}
partition sum of an isolated polymer in a good solvent,
where the excluded volume interaction is taken into account by the
requirement of self avoidance. Although details are obviously not
very physical, it should give the correct scaling behavior in the limit
of very long chains, $N\to\infty$.

It is thus not surprising that an enormous amount of effort --- both
analytical and numerical \cite{degennes,cloiz-jann} --- has been devoted
to the study of the above
model. In particular it is known that the sum in eq.(\ref{Z}) converges
up to a critical value $z_c=1/\mu$, where $Z(z)$ has a singularity
\be
   Z(z) \sim {{\rm const} \over (1-\mu z)^\gamma} \;.
\ee
Equivalently, $C_N$ obeys a scaling law $C_N \sim {\rm const}\,\mu^N
N^{\gamma-1}$.

Most of the previous works have dealt with the subcritical phase $z<z_c$,
and the supercritical case $z>z_c$ has obviously been studied much
less. In order to define the theory for $z>z_c$, one has two alternatives:
one can either make an analytic continuation, or one can regularize the
sum in eq.(\ref{Z}) by a cut-off at large $N$. The latter can be done
such that it has a very natural physical interpretation: we simply put the
chain into a finite volume $V$, and count only those configurations which
never leave the volume.

In the limit $V\to\infty$, we obtain thus a model for a single polymer
chain which fills a macroscopic volume with finite monomer density $\rho$.
In the literature this is known as the ``dense limit"
\cite{dupl-sal,schaefer}. In the following we shall study this model by
means of Monte Carlo simulations.

Actually, the dense limit can also be understood as the limit of a more
general model. There, the number of chains is fluctuating and controlled
by a second fugacity $x$, and we take the limit $x\to 0$ \cite{wheeler}.
There has been some controversy in the literature following the claim by
Gujurati \cite{gujrati} that there should be a phase transition in this
model at a finite value of $x$. It seems now clear that this claim is
wrong \cite{schaefer,wheeler}, and we shall not mention this generalization
further. In any case, the present paper will be devoted exclusively to
the case $x=0$.

All relevant scaling laws at the transition from $\rho=0$ to $\rho>0$
can be obtained from the conventional (and well established) scaling
theory for subcritical SAW's \cite{degennes}. This will be reviewed in
sec.3, where also Monte Carlo data are given in support of these
scaling laws in 3 dimensions.

In the opposite limit, when $\rho= 1$ such that the SAW visits every point
of the lattice, we are dealing with {\it hamiltonian walks}. These have
been studied in several papers, the most comprehensive study being
that of \cite{david}. They form a model for a dense polymer melt.

The intermediate region $0<\rho<1$ is particularly interesting in 2
dimensions, where it was shown in \cite{dupl-sal} that the model is a
co-dimension 0
(``self-organized" \cite{bak}) critical phenomenon. That means that various
observables are described by non-trivial scaling laws, although there
is no need for a control parameter to take any special value. In 3
dimensions there should be no such scaling laws, but we expect
logarithmic behavior instead.

It would of course be very desirable if one could verify these predictions
numerically. Some numerical results based on transfer matrix calculations
have been given in \cite{dupl-sal}. In that paper and in
\cite{tuthill,tut-glov}, also Monte Carlo simulations were presented.
But only very small systems could be simulated in \cite{dupl-sal}, and
systems with $>1$ chain (i.e., $x>0$ in the above sense) were simulated
in \cite{tuthill,tut-glov}. Thus clear-cut and significant tests of the
predictions of \cite{dupl-sal} would be most welcome. But all polymer
Monte Carlo algorithms which we are aware of become
inefficient at high densities.

It is the aim of the present paper to show that recursive implementations
\cite{random,ads3} of the incomplete enumeration \cite{redner} and of the
enrichment \cite{wall} methods can be quite efficient in simulating
polymers in the dense limit.

This is so in spite of the problem that we cannot justify our algorithms
rigorously, in the way we have to apply them in most dense limit
problems. While this seems to be mainly a formal problem in 3$d$ (we
do not encounter any serious problems unless we go to high densities),
the situation is numerically very delicate in 2$d$. As a consequence
we cannot go to really large systems and high densities, but we can go
considerably much further than previous analyses. We
will be able to check all scaling laws mentioned above with much higher
accuracy than was possible previously. We will find good agreement in
all cases except one.

The main problem in 2$d$ is that we cannot produce long chains if we
from the beginning use a fugacity $z$ which is much larger than the
critical one. A better strategy is to start building chains with $z$
very close to $z_c$, and to increase it during the growth of the chain.
Finding the optimal schedule which gives the longest chains (i.e.,
the highest densities) is very similar to optimization in a problem
with many local maxima, where greedy methods usually are sub-optimal.

In sec.2 we shall present our algorithm and the problems encountered
when applying it to dense systems. In sec.3 we shall treat the limit of
small densities, while our main results for scaling laws in 2$d$
systems will be discussed in sec.4. The very dense limit and hamiltonian
walks are treated in sec.5. The paper concludes with a
discussion in sec.6.

\section{Algorithm}

We used a recursive algorithm which can be considered as a recursive
and randomized implementation of the old enrichment method
\cite{wall,ads3}. In this algorithm we call a subroutine
STEP({\bf x},N) in order to add the next ($N$-th) step to a chain
starting at site ${\bf x_0}=0$ and ending at  ${\bf x}_N={\bf x}$.
Apart from updating whatever statistics
we want to measure, this subroutine does the following: it first marks
site {\bf x} as occupied, calls itself a random number of times at
one or more neighboring sites ${\bf x}'={\bf x}\pm {\bf e}_i$ and with
$N$ replaced by $N+1$, and marks site {\bf x} again as free before
returning to the calling routine. The first call to the subroutine from
the main program is with arguments $({\bf x}_0,1)$.

In order to be concrete, we have to specify what we mean by ``calls
itself {\it a random number of times}". Assume that the coordination
number of the lattice is ${\cal N}$, so that we can continue a
non-reversal walk without self avoidance at every step into ${\cal N}-1$
different directions. If we want to obtain a grand canonical ensemble
(i.e. a sample in which each configuration is represented in average by
${\rm const}\times z^N$ chains), we have to make in average
$P=({\cal N}-1)z$
calls during each call of STEP, and each of the ${\cal N}-1$ neighbors
must get in average the same number of calls. In incomplete enumeration
\cite{redner,random} this is achieved by taking {\it independent}
probabilities for each possible continuation, i.e. for each free
neighbor of {\bf x}. In our present implementation, in contrast,
we chose the neighbors such that the number of continuations shows
the smallest fluctuation. More precisely, we first select a random
non-reversal direction {\bf e} and call STEP at the site ${\bf
x'=x+e}$, then we select randomly another of the non-reversal directions
\footnote{In not too dense systems it is more efficient to skip the
selection of a new direction, and to use the same random ${\bf x}'$ for
all calls.}. If $P<2$, we select a random number $r\in[0,1]$ and make a
second call iff $r<P-1$. Otherwise, we choose a new random direction
(different from all previous ones) and continue. If $Q$ is the smallest
integer $\leq P$,
we try the first $Q$ directions unconditionally, and use $r$ only for
the last trial. Obviously, the above assumes that $P\leq {\cal N}-1$.
If $P={\cal N}-1$, this algorithm is equivalent to exact enumeration.

A conceptually trivial but technically very important
generalization consists in letting $P$ depend on $N$. If a sample
obtained in this way contains $n_{N,{\cal C}}$ chains with
configuration ${\cal C}$, the contribution of ${\cal C}$ to the
canonical partition sum is
\be
   {n_{N,{\cal C}}\over n_0 ({\cal N}-1)^N} \prod_{k=1}^N P_N \;.
                                            \label{n-tour}
\ee

As described here, enrichment is much more efficient than
incomplete enumeration, typically by factors 5-30 depending on the
lattice and on the density of chains. In order to understand why, we
have to understand the temporal evolution of the chain length $N$.

As pointed out in
\cite{beretti}, $N$ performs essentially a random walk with reflecting
boundary at $N=0$. If $z<z_c$, there is a bias for $N$ to decrease with
(CPU-)time, since the average number of returns from subroutine calls
is larger than the number of new calls. On the other hand, if $z>z_c$,
the walk is biased towards an increase of $N$. In either case, the
algorithm will be inefficient in creating many statistically
independent chains:
in the first case since it does not create many chains at all, in the
second because most chains will have common first parts and will thus be
strongly correlated. The most efficient algorithm is obtained when
$z$ is such that there is no net bias. Notice that the latter is true
also for dense systems. In this case, there is an additional bias
towards smaller chains which increases with $N$ and which is directly
proportional to $\rho$. In an efficient algorithm we should thus
start with $P_N\approx ({\cal N}-1)z_c$ and increase $P_N$ slowly with
$N$ in such a way that the net bias remains small.

But absence of bias is not the only criterion for efficiency. The
random walk in $N$ is also characterized
by an effective diffusion constant $D$. If $D$ is small, then $N$
changes little with time, and the algorithm is obviously inefficient.
More precisely, the CPU time needed to create one new chain and erase
it again (so that the next independent chain can be built up) is
inversely proportional to $D$. For dilute systems one can easily
convince oneself that $D = {\cal O}(1)$ for incomplete enumeration,
while it can be $>>1$ for enrichment \cite{ads3,theta,4-dim}. The
reason for the latter is easily understood. Assume that ${\cal N}>>1$
so that there is very little attrition and we can choose $P_N$
close to 1, more precisely $P_N=1+\delta$. In this case most of the
steps in the walk will be forward steps, but once we make a backward
step (since we selected an occupied site for the next subroutine call)
this step will have length $1/\delta$ in average. Since the net bias
is zero, we must thus have $1/\delta$ forward steps per each back
steps, and $D=1/\delta$. This is of course not rigorous (since the
random walk is correlated), but it gives a roughly correct estimate.

The main reason why our algorithms become inefficient at high
density (even if $P_N$ is chosen optimally, which is already not easy
to achieve) is that the above argument break down there, and the
effective value of $D$ decreases dramatically. Again this is
easily understood heuristically: in a very dense system, we have to
make many attempts for each forward step, and the algorithm will spend
much time in tempting to prolongate the chain in cases where there
simply might be no possible prolongation. Notice that this cannot
be avoided by adjusting $P_N$ so that it would be more appropriate to
each specific local configuration. $P_N$ has to be chosen uniquely for
all local configurations, and has to be such that {\it in average}
there is no bias. In this sense, polymers in the dense limit are
even worse than SAW's in random media where we can chose a different
optimized $P_N({\cal C})$ for each realization ${\cal C}$ of the
randomness \cite{random}.

In the following, we shall call ``tour" the entire random walk between
two returns to $N=0$, i.e. the set of all chains generated between two
successive returns to the main program. As described above, our
algorithm has to be used such that statistics is collected while a
tour is going on, but it must be {\it read out} only {\it after} a
tour has been completed. Only in that case eq.(\ref{n-tour}) is
correct. For instance, if we run our algorithm with $P>>({\cal
N}-1)z_c$ and stop it before finishing a tour, we will ``override"
some of the repulsive effect of self-avoidance constraint, and the
end-to-end distance will in average be smaller than the correct one
\footnote{The effective repulsion results from the fact that a
tour which involves larger end-to-end distances will last longer
and will thus contribute more to the average distance -- provided
we follow it until it is finished.}. A crucial question for the
following is whether it is possible (and if so, under what conditions)
to disregard this problem, and to stop a simulation {\it before a tour
is finished}. There are situations where this seems more or less the
only feasible strategy, since otherwise we either get large finite
size corrections (due to short chains which are included near the
end points of a tour) or extremely large CPU times.

Thus we would like to follow a strategy where we simulate essentially
a single long tour at $z>z_c$ which would not be finished at any
practically reachable time, and interrupt it after some prescribed
number of subroutine calls. It seems that this works indeed
surprisingly well in 3 dimensions (though we do not have any analytic
proof for it), but it would fail dramatically in $d=2$.

The reason for the failure in $d=2$ is easily understood. There, a
typical chain gets stuck in traps as illustrated in fig.1. If we call
$A$ the area of the trap, it takes $\sim e^{(P-P_c)A}$ steps (with
$P_c = ({\cal N}-1)z_c$) to get out of this trap again. Since the size
of the trap can be of the order of the entire lattice, the logarithm of
the time needed to get an unbiased sample is thus comparable to the
logarithm of the time needed for a complete tour.

In 3-$d$ a chain does not form traps, and we expect this problem to
be absent. Indeed, we found numerically that the chain length fluctuates
during a tour only little around an average value which is independent
of the random number sequence (with a characteristic time of the
density autocorrelation function which increases at most as a power of
the system size), provided the number of steps is
$>V/(P-P_c)$. The latter is the number of steps which would be needed
to generate a chain of length $V$ on an infinite lattice. To illustrate
the dramatic difference between the 2-$d$ and 3-$d$ cases, we show in
fig.2 histograms of chain lengths reached after $10^7$ steps. Each
histogram is based on $>100$ independent runs. In both cases, $P$ was
independent of $N$ and
$20\%$ above the critical value, and the lattices had the same
number of sites ($512\times 512$ resp. $64^3$). While we see a very
wide distribution centered at $N<<V$ for $d=2$ (panel a), the
distribution for $d=3$ (panel b) is sharply peaked at a finite density.
This density is moreover found to be independent of the lattice size.

The influence of trapping in $d=2$ can also be seen in a different way.
In fig.3 we show the average chain lengths obtained after $m=10^k$ steps
($k=3,4,5,6$ and 7) as a function of $z/z_c$ (the average {\it maximal}
length obtained during the first $m$ steps shows very similar behavior,
and is only slightly larger, as the chains hardly recede at $z\geq z_c$;
we used $z_c=0.3790524$ \cite{guttmann}). The most
conspicuous feature in fig.3 are the pronounced peaks at $z/z_c$
slightly above 1. The steep decay at $z/z_c\leq 1$ is of course trivial.
Less trivial is the fact that increasing $z$ far above $z_c$ does not
lead to longer chains. Instead it leads to more trapping. The algorithm
becomes more ``greedy" in the sense that there is more bias towards
an increase of the chain length, and it cannot so easily leave a trap
once it has run into it. In fig.3 we show results from lattices of sizes
up to $4096\times4096$ (which is practically infinite for $10^7$ steps).
In this limit of infinite lattices, the optimal
value of $z/z_c$ seems to decrease towards 1 with the number $m$ of
steps. This means that greedyness is increasingly punished with chain
length. The optimal value of $z-z_c$ scales roughly as $1/\sqrt{m}$, and
the chain lengths reached at these optimal values scale as $\sqrt{m}$.

The need for not being too greedy is very reminiscent of optimization
problems involving complex functions with many local minima.
Prototypes of such problems are provided by the traveling salesman
problem, and by the ground state search in spin glasses
\cite{mezard}. The fugacity in the present case is analogous to the
inverse temperature when treating such problems by simulated annealing
\cite{mezard}. The main difference is that in our problem we do not have
any {\it frozen} randomness. In typical hard optimization problems the
multitude of false minima is due to randomness in external conditions
which remain fixed during the optimization process. In the present case,
in contrast, it is the initial stages of the process itself which were
random and which remain ``frozen" during a {\it finite} time which
diverges however as the process goes on. In this respect our problem is
closer to the complex optimization going on in evolving biological
systems \cite{kauffman}, where it is also mainly the complexity of the
evolutionary process itself (the ``co-evolution") which leads to
multistability.

\section{Small Densities: Average Monomer Density versus Fugacity}

\subsection{Infinite Volume Limit}

The standard scaling theory for free SAW's \cite{degennes,cloiz-jann}
describes the behavior of SAW's in the grand canonical ensemble in the
limit $z\to z_c-\epsilon$, i.e. if we approach the critical fugacity
from the subcritical region. In the present section we propose that
this scaling behavior can be extended slightly above $z_c$. Thus we
should be able to predict also the
behavior of dense SAW's in the limit of small densities in terms of the
usual critical exponents and scaling functions for non-dense SAW's.

Let us consider the density a finite box of volume $V=L^d$ with periodic
boundary conditions. Let $\rho({\bf x},z)$ be the density of monomers
in an ensemble where a single polymer is attached with one end to the
origin ${\bf x}=0$. The length of the monomer is determined implicitly
by fixing the fugacity $z$, i.e. by giving each $N$-step chain a weight
$\propto z^N$. For
\be
   \epsilon \equiv {z-z_c\over z_c} < 0
\ee
this density becomes independent
of $L$ for $L\to\infty$, and obeys the scaling law \cite{degennes}
\be
   \rho({\bf x},z) \approx r^{1/\nu -d} g(r^{1/\nu}\epsilon) \;,\quad
             r=|{\bf x}| \;,                     \label{rho}
\ee
where $g(\zeta)$ is a universal scaling function which can be assumed to
be analytic at $\zeta=0$. Notice that the latter is a direct consequence
of our choice of scaling variable. More conventional would be the choice
$r\epsilon^\nu$, but it is clear that this would generate a singularity
in $\rho({\bf x},z)$ at $z=z_c$ if $g$ were analytic.

We now assume that eq.(\ref{rho}) holds also for $\epsilon>0$. In this
case the polymer should fill a large volume with constant density (except
for a region near the origin), and thus $\rho({\bf x},z)\to {\rm const}$
for ${\bf x}\to\infty$. This is only possible if
\be
   g(\zeta)\sim \zeta^{d\nu-1} \;,
\ee
and from this follows \cite{saleur}
\be
   \rho_\infty(z) = \lim_{{\bf x}\to \infty} \rho({\bf x},z)
         \sim \epsilon^{d\nu-1}\;.      \label{rho-scal}
\ee

In 3 dimensions we had no difficulty in checking this directly. We used
systems of different sizes for different values of $z$, but for all
$z$ they were large enough so that the fluctuations in chain length
were much smaller than the average chain lengths. This is necessary for
the average density to be independent of the box size. Otherwise, we would
have had a non-negligible chance that the chain length goes occasionally
down to zero, which would obviously depend on box sizes. Specifically,
this meant for the runs closest to $z_c$ that we used box
sizes $512^3$, and allowed chain lengths up to $3.3\times 10^5$. Notice
that this implied also that we had to take statistics without waiting
for tours to become completed, as discussed in the last section.

Results from these runs are shown in fig.4. In this plot we used
$z_c=0.2134908$ which is the value found in \cite{ads3} (with error
$\pm 0.0000005$) for SAW's attached to a planar surface. This value is
slightly smaller than the previous best value by one of us \cite{grass}
for free SAW's in infinite volume, but additional unpublished simulations
with the latter geometry gave a value in
perfect agreement with the above. In contrast, a much larger value
with comparable accuracy ($0.2134987\pm 0.0000010$) was found from
exact enumerations in \cite{macd}. We see from fig.4 that we can again
exclude the latter. Fig.4 is obviously compatible with scaling. A least
square fit would give an exponent $d\nu-1 = 0.728\pm 0.1$ or
$\nu=0.579\pm 0.003$, slightly below the best current estimate
$\nu=0.587$ \cite{nickel}. Indeed, a careful look at fig.4 shows a
slight downward curvature which might explain this small discrepancy.

Before going on we should point out that eq.(\ref{rho-scal}) has
an important consequence for the free energy per monomer, as was
pointed out in \cite{dupl-sal}. If we assume that the free energy
is an extensive quantity in the thermodynamic limit, we should have
\be
   {1\over N} \ln C_{N,L} \approx -f(N/L^d)      \label{CNLinf}
\ee
with $f(\rho)\to -\ln \mu$ for $\rho\to 0$. If the grand canonical
partition sum
\be
   Z_L(z) = \sum_N z^NC_{N,L}\approx \int dN\,z^N e^{-N f(N/L^d)}
\ee
is estimated by the saddle point approximation, we find the saddle point at
\be
   {\partial \over \partial\rho}[\rho f(\rho)] = \ln z \;.
\ee
Putting $f(\rho)=-\ln \mu+\Delta f(\rho)=\ln z_c+\Delta f(\rho)$,
we obtain
\be
  - {\partial \over \partial\rho}[\rho \Delta f(\rho)] =
             \ln(z_c/z) \sim \rho^{1/(d\nu-1)} \;.
\ee
Integrating this gives \cite{dupl-sal}
\be
   f(\rho) = -\ln \mu + {\rm const}\, \rho^{1/(d\nu-1)} \;. \label{df}
\ee

\subsection{Finite Volume Corrections}

In $d=2$ we cannot proceed in the same straightforward way for the reasons
explained in sec.2. Instead, we must work on lattices which are
small enough, and with fugacities which are close enough to $z_c$,
so that all tours finish. Indeed, in order not to introduce any bias
we have to prescribe the number of tours to be made, not the CPU time.

This means however that we are never really in the limit described by
eq.(\ref{rho-scal}), and we have to consider finite-$L$ corrections
to it. Assuming that there is a unique diverging length scale
$\xi\sim N^\nu\sim \epsilon^{-\nu}$ in the polymer problem, any
dependence on $L$ can only enter through the dimensionless ratio
$\xi/L$, and we can make the ansatz
\be
   \rho_L(z) = \langle N\rangle/L^d \approx \epsilon^{d\nu-1}
                  \phi(\epsilon L^{1/\nu})              \label{rhoL}
\ee
where $\phi(\zeta)$ is a new scaling function with $\phi(0)$ finite.
Notice that this is only an ansatz for the spatially averaged density,
and it implies no assumption about its ${\bf x}$-dependence.

Alternatively we can start with the canonical ensemble, i.e. with chains
of fixed length $N$. Generalizing the usual ansatz for isolated chains
in infinite volume and the discussion at the end of the last subsection,
we assume
\be
   C_{N,L} \approx N^{\gamma-1} e^{-Nf(N/L^d)} \;.     \label{CNL}
\ee
In contrast to eq.(\ref{rhoL}), the last ansatz should hold also for
large densities, and even in the hamiltonian limit $\rho=1$. In the
latter, it is known that \cite{dupl-sal}
\be
   C_{N,N}  \approx N^{\gamma_H-1} [\mu_H]^N \;,       \label{manh}
\ee
where $\gamma_H$ depends not only on the dimensionality $d$ but also on
the boundary conditions. In $d=2$ it has been conjectured that $\gamma_H$
for all lattices is the same as for the Manhattan lattice, where it has
the value 35/16 for periodic boundary conditions \cite{dupl-sal}. In that
reference it is also conjectured that {\it all} the power behavior in
eq.(\ref{manh}) is for $N<L^2$ transformed into a term $L^{2(\gamma_H-1)}$.
The latter would be in conflict with eq.(\ref{CNL}), whence we believe
that it is just a trivial mistake.

It is straightforward to show that eqs.(\ref{rhoL}) and (\ref{CNL})
are mutually consistent for any value of $\gamma>0$, provided just that
$f(\rho)$ satisfies eq.(\ref{df}). Indeed, with eq.(\ref{df}) the
ansatz eq.(\ref{CNL}) becomes $C_{N,L} \approx N^{\gamma-1} \mu^N
\exp(-{\rm const}N\rho^{1/(d\nu-1)}) =  N^{\gamma-1} \mu^N \chi(N^\nu/L)$.
Inserting this into the definition of $\langle N\rangle$ and replacing
the summation over $N$ by an integral gives eq.(\ref{rhoL}).

In order to verify eqs.(\ref{CNL}) and (\ref{df}) in 2 dimensions,
we made simulations on square lattices of sizes between $16\times 16$
and $256\times 256$. During the simulations $z$ was not kept fixed
but depended on $N$ (in order to reach longer chains), but as pointed
out in sec.2 it is trivial to compute $C_{N,L}$ from this. In fig.5
we plotted $\mu(L) = e^{-f(\rho)}$ as defined through eq.(\ref{CNL}).
Actually, to reduce finite size corrections we replaced the argument of
$f$ by $(N-L)/L^2$. This is suggested by the fact that the finiteness of
the lattice cannot be felt by chains with $N<L$ (we used periodic
boundary conditions). We see reasonable scaling, i.e. the functions
$f$ defined in this way really depend only on the chain density. We
should stress that this would not be so if we had omitted the first
factor in eq.(\ref{CNL}). To verify eq.(\ref{df}) we plotted $\Delta
f$ in fig.6 against $z-z_c$. We see that the prediction $\Delta f
\sim \rho^2$ is satisfied reasonably well. The agreement is certainly
far from perfect, but it improves with increasing $L$, and it would be
much worse if we would not have taken into account the first factor in
eq.(\ref{CNL}).

\section{Spatial Properties in Dense 2-D Systems}

Some of the most striking predictions concern the
spatial structure of dense SAW's in 2 dimensions. In spite of having
no obvious control parameter which is set to any special value, the
system is predicted to be critical, and several correlation
functions should obey non-trivial anomalous scaling laws. We will not
attempt here to rederive them but simply refer to \cite{dupl-sal} for
derivations.

\subsection{Grand Canonical End-to-End Distance Distribution}

The first saling law which we want to discuss here concerns the probability
that a SAW has an end-to-end distance $R$, irrespective of the number
of steps $N$ it involves. Due to the effective repulsion provided by the
self avoidance constraint, one should expect that this {\it increases}
for small $R$. In the spin model analogy this means that the spin-spin
correlation increases with distance, a paradoxical result which was
put in question in \cite{gujrati2}.

More precisely, we studied chains starting in the center of a circular
disk of radius $r$ cut out from a square lattice. No chain is allowed to
reach further than $r$. For $R<<r$, the prediction by Nienhuis \cite{nien}
is
\be
   P(R;z) \sim R^{3/8}\;.                  \label{nienh}
\ee
In order to check this, we show in fig.7 results from a disk with radius
$r=45$. The fugacity was chosen as $z=0.39$, i.e. about 3\% larger than
the critical value. The data are normalized by dividing them by the
number of chains with $R=1$ (which is also the number of self avoiding
rings). We see in fig.7 large fluctuations at low values of $R$ which
are {\it not statistical}, but due to the discreteness of the lattice.
Statistical errors are much smaller. In the intermediate range of $R$
we indeed observe the predicted scaling regime. The drop at very large
$R$ is simple due to the finiteness of the lattice \footnote{the choice
of a circular boundary was mainly to keep this drop confined to a
narrow region in $R$.}.

Analogous data for three dimensions are shown in fig.8. There we show
two curves, one for a moderately large spheric volume ($r=40$) but
reasonably large fugacity ($z=0.215=1.0071 z_c$), the other for a much
larger volume ($r=77$) but much closer to the critical point ($z=.214
= 1.0023z_c$). Both curves show that $P(R;z)$ increases for small $R$,
but the increase is much slower than for $d=2$ and does not seem to
follow a power law.

\subsection{Number of Loops vs. Number of Open Chains}

Instead of working at fixed fugacity, let us now consider the number
of SAW's and of self avoiding loops at fixed densities. Let us call
the latter $C_{N,L}^{(0)}$. We work on square lattices with periodic
boundary conditions, and count only loops which pass through the
origin.  According to \cite{dupl-sal}, we expect
\be
   {C_{N,L}^{(0)}\over C_{N,L}} \sim L^{-2\gamma_D}\;,\qquad \gamma_D
          = 19/16\;.                                  \label{loops}
\ee
Numerically, this was checked in \cite{dupl-sal} for $L\leq 16$, at
density $N/L^2 = 0.4$. Actually, in general $0.4L^2$ is not integer.
Thus for each $L$ the chain length $N$ nearest to $0.4L^2$ had to
be taken, and the density slightly fluctuated around 0.4.
While the global behavior was indeed as predicted, it was
superimposed by strong odd/even oscillations.

We found that this odd/even effect is mainly due to the following:
due to the periodic boundary conditions our lattice is indeed a 2-$d$
torus. Loops can either wind around the torus or have trivial topology
(i.e., be homotopic to a point). The latter loops and those which
wind twice around the torus always have $N$ even. Walks which wind
once around the torus have $N$ even on lattices with even $L$, and
vice versa. The total number of of loops depends smoothly on $L$.
Thus the number of even-$N$ chains at even $L$ are larger than
the number of chains at $L\pm 1$ at any near-by value of
$N$. In order to eliminate this effect, we thus plot in fig.9 not
precisely the ratio $C_{N,L}^{(0)}/C_{N,L}$ but rather
$(C_{N-1,L}^{(0)}+2C_{N,L}^{(0)}+ C_{N+1,L}^{(0)}) /2C_{N,L}$. We
see indeed a perfectly straight line on a log-log plot up to $L\approx
40$ (the fluctuations for even larger $L$ are statistical). The
theoretical prediction is indicated in fig.9 by a straight
line, and is obviously in perfect agreement with our data.

Recently there has been some discussion about the value of $\gamma_D$.
Based on exact enumerations of collapsed self-attracting SAW's on
unbounded lattices, Bennett-Wood {\it et al.} \cite{bennett} found
$\gamma_D=0.92\pm 0.09$, while exact solution for the honeycomb lattice
with strip geometry gave $\gamma_D=1$ \cite{batch}. Our data clearly
disfavor these values (our estimate is $\gamma_D=1.14 \pm 0.05$). One
reason of the discrepancy with the work of \cite{bennett} might be
that collapsed self-attracting SAW's at finite
temperatures are not equivalent to dense SAW's, in contrast to common
belief \cite{owcz-reply}. Another reason might be that $\gamma_D$
depends on the boundary condition, again in contrast to common belief.

\subsection{Number of Chains, Absolute Values}

As a last scaling law, we used the same data to check the prediction of
\cite{dupl-sal} that the number of self avoiding chains at fixed density
$0<\rho<1$ scales with a $\gamma$-exponent equal to $\gamma_D+1$, i.e. as
\be
   C_{\rho L^2,L} \sim L^{2\gamma_D}e^{-L^2\rho f(\rho)}   \label{loop1}
\ee
(compare eq.(\ref{manh}). This would imply that the number of
loops scales without any power prefactor. It agrees with the
behavior in the hamiltonian limit, but we pointed already out in
sec.3 that it cannot be correct in the limit $\rho\to 0$. In the
present subsection we shall test whether it holds for intermediate
values of $\rho$, more precisely at $\rho=0.4$. To get rid of the
unknown constant prefactor in eq.(\ref{loop1}), we first form the ratio
\be
   C_{N(2L),2L}/ C_{N(L),L} \;.   \label{X}
\ee
Here $N(L)$ is defined as the integer closest to $0.4L^2$.
We then define an effective free energy per monomer as
\be
   f_L(\rho) = -{\ln[C_{N(2L),2L}/ C_{N(L),L}] \over N(2L)-N(L)}
                                               \label{f}
\ee
and plot $f_L(\rho)$ against $1/L^2$. If eq.(\ref{loop1}) is correct,
this should converge for $L\to \infty$ towards $f(\rho)$ with slope
$[\gamma_D \ln 4]/[3\rho]$. Otherwise, if eq.(\ref{CNL}) is correct,
the slope at $1/L^2=0$ should be $[(\gamma-1)\ln 4]/[3\rho]$. This
plot is shown in fig.10. It suggests that eq.(\ref{loop1}) is not
correct, and eq.(\ref{CNL}) is preferred. A more precise
statement about the value of $\gamma$ is difficult because of
the large fluctuations visible in fig.10. They are partly due to the
fact that $\rho$ is not strictly constant, but there seems also to be
some large scale curvature superimposed. In view of this, our estimate
from fig.10 is $\gamma = 1.25\pm .10$. This could mean that $\gamma$
indeed has its critical value 43/32, and that eq.(\ref{CNL}) holds for
all finite densities.

\section{Very Dense Systems and Hamiltonian Limit}

It is clear from the above that our method becomes less and less
efficient if we go to higher and higher densities. Thus it is clear
that we can study hamiltonian chains only for very small systems.

In fig.11 we show results for 2-$d$ (panel a) and 3-$d$ (panel b) systems
with free boundary conditions. Due to the boundary conditions, all
statistics depend on the starting point of the chains. In fig.11 we show
average values, obtained by averaging over all $L^d$ starting points on
a square (resp. cubic) lattice. The observable shown is the decadic
logarithm of the number of chains. Our results are in good agreement
with expectations \cite{dupl-sal,david} that
\be
    \log\langle C_{N,L}\rangle = \log [L^{-d}
                         \sum_{\mbox{\scriptsize starting pos.}}
    C_{N,L} ]  \sim -Nf(N/L^d) -B(N/L^d)L^{d-1}
\ee
up to logarithmic terms.

The lattice sizes shown in fig.11 are rather modest: $L\leq 10$ for
$d=2$, and $L\leq 5$ for $d=3$. Indeed, we tried larger lattices, but we
were simply unable to generate even a single hamiltonian chain on any
larger lattice with our method. While we estimated that $L=6$ for $d=3$
and $L=11$ for $d=2$ should be feasible with unduly large amounts of
CPU time, lattices beyond that should be unaccessible with (variants of)
our algorithm, and with present day computers.

The data shown in fig.11 were obtained by splitting the range $[1,
L^d-1]$ of chain lengths into 3 subintervals $[1,N_1]$, $[N_1+1,N_2]$
and $[N_2+1,L^d-1]$. On the middle interval we performed the Monte
Carlo algorithm as described in sec.2, while we made {\it complete
enumerations} of all chains in the first and third intervals (typically,
$N_1=8$ for $d=2$ and $N_1=6$ for $N=3$, while $N_2\approx 0.75 L^d$).
The exact enumeration on the first interval slightly reduced CPU time
and statistical fluctuations, but was not very important for efficiency.
In contrast, performing exact enumeration for $N\geq N_2$ was crucial.
It implied that we found {\it all} extensions of random samples of
chains with length $N_2$.

Indeed, as seen in fig.11, the number of chains decreases dramatically
when approaching $N\to L^d$. For $d=2$ and $L=10$, e.g., we see that
the number of hamiltonian chains is less than the number of chains with
$N\approx 75$ by more than 7 orders of magnitude! If we try to find
hamiltonian chains by extending randomly chosen self avoiding chains
which fill the lattice partially, we need thus $>10^7$ chains of length
75 in order to have just a single success (the data shown in fig.11a
are based on $6\times 10^8$ chains of length 75, those of fig.11b ---
where this factor is $>10^8$ --- are based on $1.4\times 10^9$ chains).

As seen from fig.11, this decrease of $C_{N,L}$ when approaching the
hamiltonian limit becomes rapidly stronger as we increase $L$, rendering
our algorithm impracticable on larger lattices. Things are very similar
for periodic boundary conditions. The situation is slightly better if we
select the chains not randomly but such that dense chains are favored
from the very beginning. Thus, the success rate is slightly higher if
we use the Rosenbluth method \cite{rosen}, but not
enough to treat lattices larger than $11\times 11$ resp. $6^3$.

We should stress that this represents a fundamental obstacle for
{\it any} algorithm which attempts to generate hamiltonian walks by
completing randomly chosen chains which partially fill the lattice. Just
such an algorithm was presented in \cite{siep}. In that paper first a
non-random hamiltonian chain was formed (this is always easy). This
chain is cut into two pieces at a randomly chosen site. The shorter
half is erased, and completion of the longer part to a new hamiltonian
chain is attempted using the Rosenbluth method. If this attempt is
successful, the old chain is replaced, while it is restored in the
case of failure. After that, a new random cutting site is chosen and
the process is repeated. It was claimed in \cite{siep} that the
algorithm was successful in generating random hamiltonian chains on
2-$d$ lattices with up to $21\times 21$ sites and with periodic bc.
According to our estimates this seems very unlikely. We conjecture
that all successful attempts on the largest lattices in
\cite{siep} involved cutting the chain
near one of its ends. The chance for completing it when cutting near
the center should have been practically zero, and thus the central
piece of the chain should not have been changed at all.

In order to improve the situation, one can use a sample of partial
chains which is strongly biased in favor of chains which do have at
least one hamiltonian extension. We created such samples by growing
random chains of length $N\approx L^d/2$ and discarding all chains
for which the set of yet unvisited sites is disconnected. On square
lattices with $L=11$, e.g., this reduced a sample of chains with
$N=60$ by a factor $\approx 10^3$. This rendered feasible complete
enumeration of all extensions of the remaining chains. Repeating
the test for connectivity of the unvisited sites several times
improved further the efficiency. In this way we could generate a few
hamiltonian chains with $L=12$ ($d=2$) resp. $L=6$ ($d=3$), but on
larger lattices the method failed again.

Two surprising results were found in \cite{siep}: first of all,
$\langle R_N^2\rangle $ increased faster than linearly with
$N$, $\langle R_N^2\rangle \sim N^{1.26}$, and secondly there were
strong odd/even oscillations superimposed on that growth. In fig. 12
we show our data for $\langle R_N^2\rangle $, also for hamiltonian
chains on square lattices with periodic boundary conditions. We see
no indication for a faster than linear growth. We do see some odd/even
oscillations, but the amplitude is much weaker than that observed in
\cite{siep}, and has {\it opposite sign}.

\section{Discussion}

In this paper we have presented simulations of single chains in the dense
limit. We have seen that such simulations are very easy in $\geq 3$
dimensions when using a recursive and stochastic implementation of the
enrichment method, unless one wants to go to very large densities.
More interesting is the 2-$d$ problem due do
universal anomalous scaling laws which should hold there. We were
indeed able to test these, with much higher precision than in previous
analyses, although our method is much less efficient in 2$d$ than in
3$d$. The reason for the latter is that our algorithm lets SAW's
run into `cages'. In the dense limit we have to use supercritical
fugacity, implying that it becomes hard to leave a cage. This becomes
the more pronounced the more the fugacity is above its critical value.
This is very reminiscent of `greedy' optimization algorithms in
problems with frozen randomness.

In spite of this problem we were able to perform simulations of much
longer chains than previously. With one minor exception we verified
all scaling laws predicted for dense 2-$d$ polymers. This proves that
this system is indeed a generic (co-dimension 1) critical phenomenon.

We also performed simulations of the hamiltonian limit, and showed
that neither our algorithm nor any variant thereof can be efficient.
If one wants to create a random sample of hamiltonian chains by
completing partially filling chains, one has to use for the latter a
sample which is strongly biased in favor of such chains which have
at least one hamiltonian extension.

In the present paper we have reported only on simulations of a-thermal
polymer chains. In \cite{theta}, we have simulated dense polymers near
the $\theta$-point. These simulations were essential in eliminating
surface effects in the coil-globule transition, and in understanding
the nature of this transition. For more details we refer to
\cite{theta}.

Acknowledgement: We are indebted to Drs. Frenkel and Siepmann for
discussions about their work. The present work was supported by
DFG through SFB 237 and Gradu\-iertenkolleg `Numerische und
Feldtheoretische Methoden in der Hochenergiephysik und Statistischen
Mechanik'.

\eject

\vspace{1.5cm}

\eject

\section*{Figure Captions:}

{\bf Fig.1:}  Trapped 2-$d$ SAW. In supercritical simulations using the
present algorithm, there is a bias towards increase of the chain length,
and it can take extremely long until such a trap is left again.

{\bf Fig.2:} Histograms of chain lengths reached after $10^7$ subroutine
calls in supercritical simulations. Panel a: 2-$d$ lattice of $512\times
512$ sites, $\mu=0.454863$; panel b: 3-$d$ lattice of size $64^3$, $\mu=
0.256189$.

{\bf Fig.3:} Average chain lengths obtained after $10^k$ subroutine
calls ($k=3,\ldots 7$) on 2-$d$ lattices of sizes up to $4096\times
4096$, plotted against $z/z_c$.
For each value of $z$, data are averages over $500$ independent runs.

{\bf Fig.4:} Log-log plot of the density of supercritical SAW's filling
a finite but large 3-$d$ volume, in dependence of the fugacity $\mu$.
The critical fugacity was chosen according to \cite{ads3}.

{\bf Fig.5:} Effective coordination numbers per monomer for 2-$d$ chains,
after dividing out the logarithmic corrections from $C_{N,L}$ (see
eq.(\ref{CNL}). On the abscissa is plotted the corrected density
$(N-L)/L^d$.

{\bf Fig.6:} Log-log plot of $\Delta f$ against the corrected density,
obtained from the same data as in fig.5.

{\bf Fig.7:} Log-log plot of the number of chains with end-to-end
distance $R$ plotted against $R^2$. These data were obtained on a
disk of radius $r=45$ at fixed fugacity $z=0.39$. They are normalized
to 1 at $R=1$. The straight
line has slope 3/16, and indicates the theoretical prediction.

{\bf Fig.8:} Simular to fig.7 but for three dimensions.

{\bf Fig.9:} Number of loops divided by number of chains, both at fixed
density $\rho=0.4$, plotted against $L$. The theoretical prediction is
indicated by the straight line with slope -19/8.

{\bf Fig.10:} Free energy per monomer $f_L(\rho=0.4)$ against $1/L^2$.
Indeed, since $0.4L^2$ is generally not an
integer, the density slightly oscillates around the value 0.4. This is
mainly responsible for the fluctuations seen in the data, as statistical
errors are comparable to the size of the dots.

{\bf Fig.11:} Decadic logarithms of the number of chains on lattices
with free boundaries, averaged over the position of the starting point.
For $d=2$ (panel a), curves are for $L=4,\ldots 10$, while they
correspond to $L=3,4$ and 5 for $d=3$ (panel b).

{\bf Fig.12:} $\langle R^2\rangle^{1/2}$ against $L$ for hamiltonian
chains on square lattices with periodic boundary conditions. The data
for $L\leq 6$ are based on exact enumerations and thus free of
statistical errors.

\eject

\end{document}